\documentclass[usenatbib]{mn2e}

\usepackage{epsf}
\usepackage{graphicx}

\newcommand{\alphah}{\alpha_{\rm h}}
\newcommand{\nbf}{ }
\newcommand{\sigmah}{\sigma_{\rm h}}
\newcommand{\apj}{ApJ}
\newcommand{\apjs}{ApJS}

\newcommand{\nat}{Nature}
\newcommand{\mnras}{MNRAS}
\newcommand{\Msol}{M$_{\odot}$}
\newcommand{\lsim}{\mathrel{\hbox{\rlap{\hbox{\lower4pt\hbox{$\sim$}}}\hbox{$<$}}}}

\begin{document}

\title[Merger Trees]{Generating Dark Matter Halo Merger Trees}
\author[Parkinson et al.] {
\parbox{\textwidth}{Hannah Parkinson, Shaun Cole, John Helly}
\vspace*{4pt} \\
Institute of Computational Cosmology, Department of Physics, 
University of Durham, South Road, Durham DH1 3LE, UK 
}
\maketitle

\begin{abstract}
    We present a new Monte-Carlo algorithm to generate merger trees
    describing the formation history of dark matter halos.  The
    algorithm is a modification of the algorithm of \cite{cole00} used
    in the GALFORM semi-analytic galaxy formation model.  As such, it
    is based on the Extended Press-Schechter theory and so should be
    applicable to hierarchical models with a wide range of power
    spectra and cosmological models.  It is tuned to be in
    accurate agreement with the conditional mass functions found in
    the analysis of merger trees extracted from the $\Lambda$CDM
    Millennium N-body simulation. We present a comparison of its
    predictions not only with these conditional mass functions, but
    also with additional statistics of the Millennium Simulation halo
    merger histories.  In all cases we find it to be in good agreement
    with the Millennium Simulation and thus it should prove to be a
    very useful tool for semi-analytic models of galaxy formation and
    for modelling hierarchical structure formation in general.  We
    have made our merger tree generation code and code to navigate the
    trees available at
    http://star-www.dur.ac.uk/$\tilde{\hphantom{n}}$cole/merger\_trees
    .
\end{abstract}

\section{Introduction}
\label{sec:intro}

In hierarchical models of structure formation, such as $\Lambda$CDM,
the formation of a dark matter (DM) halo through accretion and repeated
mergers can be described by a merger tree \citep{lc93}. The merger
trees, which list the progenitors of a given halo at a series of
redshifts and describe the sequence in which they merge together,
contain essentially all the information one needs about the DM when building
models of the other processes involved in galaxy formation. Thus, the
merger trees, whether extracted from N-body simulations such as the
Millennium Simulation \citep{springel05} or generated by Monte-Carlo (MC)
algorithms \citep[e.g.][]{sheth99,somerville99,cole00}, 
provide the framework within which
one can model the additional astrophysical processes of galaxy
formation \citep{cole00,kauffmann93,somerville99}.

The statistical properties of Monte-Carlo merger trees based on the
approximate Extended Press-Schechter (EPS) theory
\citep{bcek,bower91,lc93} are not in perfect agreement with those
built from high resolution, highly non-linear N-body simulations
\citep[e.g.][]{helly07}. We have found that when the same
semi-analytic galaxy formation model is run first with MC trees and
then with N-body trees there can be significant differences in the
properties of their resulting galaxy populations. In some ways these
differences are minor as small changes in the uncertain parameters of
the star formation and feedback prescriptions can often bring them
back into alignment. However, it would be far better if MC and N-body
merger trees were in much better agreement. For instance, this would
allow galaxy formation models to be run first on MC trees and
parameters
including cosmological parameters tuned to match observed galaxy
properties
in advance of running an expensive N-body simulation which will
furnish the positional information needed to make galaxy clustering
predictions.
Additionally, as N-body simulations always have poorer mass
resolution than can be obtained with MC merger trees, one would like
to be able to use MC trees of varying resolution to assess the impact
of the limited resolution of the N-body simulation. This can be hard to
achieve when the two sets of trees differ systematically. 

In this paper we present a modification to the MC merger tree
algorithm of \citet{cole00} that we tune to be in accurate agreement
with the statistical properties of the Millennium Simulation merger
trees that were presented in \citet{helly07}.
In Section~\ref{sec:MC} we describe both the original EPS MC algorithm
as implemented in \cite{cole00} and our modification.
Section~\ref{sec:compare} compares the results of our modified
algorithm with the original and with the statistics of merger trees
from the Millennium Simulation. 
We briefly discuss relationship of our algorithm to other
models in Section~\ref{sec:disc}  and 
conclude in Section~\ref{sec:conc}.

\section{The Monte-Carlo Algorithm}
\label{sec:MC}

In the following sections, we briefly review the Monte-Carlo
algorithm implemented in the GALFORM semi-analytic code \citep{cole00} 
and then describe how we modify it to achieve more accurate
agreement with simulation data.

\subsection{The GALFORM Algorithm}
\label{sec:GALFORM}

The merger tree algorithm employed in the GALFORM semi-analytic
model uses as its starting point the conditional mass function
\begin{eqnarray}
\lefteqn{\displaystyle{
f(M_1 \vert M_2)\, d\ln M_1 = 
\sqrt{\frac{2}{\pi}} \,
\frac{\sigma_1^2
  (\delta_1-\delta_2)}{[\sigma_1^2-\sigma_2^2]^{3/2}} } \, \times} &&
\nonumber \\
&&\displaystyle{ \exp\left[ - \frac{1}{2} \frac{(\delta_1-\delta_2)^2}{(\sigma_1^2-\sigma_2^2)}\right]
\left\vert \frac{d\ln\sigma}{d\ln M_1} \right\vert \,
d\ln M_1} ,
\label{eq:cmff}
\end{eqnarray}
given by extended Press-Schechter theory \citep{bcek,bower91,lc93}.
Here $f(M_1 \vert M_2)$ represents the fraction of mass from halos of
mass $M_2$ at redshift $z_2$ that is contained in progenitor halos of
mass $M_1$ at an earlier redshift $z_1$. The linear density thresholds
for collapse at these two redshifts are $\delta_1$ and~$\delta_2$
\citep[e.g.][]{ecf96} . The rms linear density fluctuation
extrapolated to $z=0$ in spheres containing mass $M$ is denoted
$\sigma(M)$ with $\sigma_1\equiv\sigma(M_1)$ and
$\sigma_2\equiv\sigma(M_2)$. Taking the limit of $f(M_1 \vert M_2)$ as
$z_1 \rightarrow z_2$ one finds,
\begin{eqnarray}
\lefteqn{\displaystyle{
  \frac{df}{dz_1}\Big\vert_{z_1=z_2} d \ln M_1 \, dz_1 = }}&& \nonumber \\
&&\sqrt\frac{2}{\pi}  
\frac{\sigma_1^2}{[\sigma_1^2-\sigma_2^2]^{3/2}}\ \frac{d\delta_1}{dz_1} 
\Big\vert\frac{d \ln \sigma_1}{d\ln M_1}\Big\vert\ d \ln M_1 \ dz_1, 
\label{eq:lin}
\end{eqnarray} 
which implies that the mean number of halos of mass $M_1$ into which
a halo of mass $M_2$ splits when one takes a step 
$dz_1$ up in redshift is 
\begin{equation}
\frac{dN}{d M_1} =  \frac{1}{M_1} \ \frac{df}{dz_1} \frac{M_2}{M_1} dz_1
\qquad (M_1 < M_2) .
\label{eq:dnp}
\end{equation}
Then, by specifying a required mass resolution, $M_{\rm res}$, for the
algorithm on can integrate to determine 
\begin{equation}
P=\int_{M_{\rm res}}^{M_2/2} \frac{dN}{d M_1} \ dM_1,
\label{eq:P}
\end{equation}
which is the mean number of progenitors with masses $M_1$ in the
interval $M_{\rm res}<M_1<M_2/2$
and
\begin{equation}
F= \int_{0}^{M_{\rm res}} \frac{dN}{d M_1}\ \frac{M_1}{M_2} \
dM_1,
\label{eq:F}
\end{equation}
which is the fraction of mass of the final object in progenitors below
this resolution limit. Note that both these quantities are proportional 
to the redshift step, $dz_1$, by virtue of equation~(\ref{eq:dnp})

The GALFORM merger tree algorithm then proceeds as follows.  Firstly,
choose a mass and redshift, $z$, for the final halo in the merger
tree. Then, pick a redshift step, $dz_1$, such that $P \ll 1$, to
ensure that the halo is unlikely to have more than two progenitors at
the earlier redshift $z+dz$.  Next, generate a uniform random number,
$R$, in the interval $0$ to $1$. If $R>P$, then the main halo is not
split at this step. We simply reduce its mass to $M_2(1 - F)$ to
account for mass accreted in unresolved halos. Alternatively if $R \le
P$, then we generate a random value of $M_1$ in the range $M_{\rm res}
> M_1 >M_2 /2$, consistent with the distribution given by
equation~(\ref{eq:dnp}), to produce two new halos with masses $M_1$
and $M_2(1-F) - M_1$.  The same process is repeated on each new halo
at successive redshift steps to build up a complete tree. More details
are given in Appendix~A.

\subsection{The Modified Algorithm}
\label{sec:MA}

{\nbf 
The binary merger algorithm described above  fully respects a natural
symmetry that whenever one fragment
has mass $M_1$ the other must have mass $M_2-M_1$
(at least in the limit of $M_{\rm res} \rightarrow 0$). 
This means that it is not consistent with 
EPS theory as equation~(\ref{eq:dnp}) does not satisfy this symmetry
of remaining unchanged when $M_1\rightarrow M_2-M_1$
\citep{lc93,benson05}. To force the required symmetry the algorithm
only uses equation~(\ref{eq:dnp}) for $M_1<M_2/2$ and ignores its
predictions for $M_1>M_2/2$.
The algorithm is also unsatisfactory because the EPS 
conditional mass functions, and also the original Press-Schechter 
mass function, do not accurately match what is found in N-body simulations
 \citep{st99,jenkins01,helly07}}.
However, many statistical properties of the merger trees produced by
the above algorithm have trends with mass and redshift that agrees well
with those of merger trees constructed from high resolution N-body
simulations, but with increasing redshift they systematically
underestimate the mass of the most massive progenitors
\citep{helly07}. (In practice, the problem is ameliorated in GALFORM
by starting the merger tree construction at higher redshift.)
Here, our aim is to reduce these systematic
differences. Given the simplicity and zeroth order success of the
original GALFORM algorithm, it seems reasonable to try modifying it by
perturbing the basic function that drives the algorithm. Namely we
consider replacing the function defined in equation~(\ref{eq:dnp}) by
making the substitution
\begin{equation}
  \frac{dN}{dM_1} \rightarrow  \frac{dN}{dM_1} \ 
G(\sigma_1/\sigma_2,\delta_2/\sigma_2) .
\label{eq:dnp_mod}
\end{equation}
Here $G(\sigma_1/\sigma_2,\delta_2/\sigma_2)$ is the ``perturbing''
function which we expect to be of order unity for most of the range of
interest. The choice that the function $G$ should only depend on the
ratios $\sigma_1/\sigma_2$ and $\delta_2/\sigma_2$ is motivated by the
desire that the algorithm should preserve self-similarity if used in a
flat $\Omega_m=1$ cosmology with scale free initial
conditions \cite[e.g. see][]{efstathiou88}. The dependence on
$\delta_2/\sigma_2$ allows the halo splitting rate to be modified as a
function of $M_2/M_*$, where the characteristic non-linear mass, $M_*$, is defined
by $\sigma(M_*)=\delta$, while the dependence on $\sigma_1/\sigma_2$
allows the mass distribution of the resulting fragments to be modified.
Restricting the dependence of the function to only these parameters
is necessary to preserve self-similarity, but on its own does not
guarantee self-similarity.
The additional unwanted freedom we hope to remove by fitting to the
statistical properties of the Millennium Simulation merger trees, as
presented in \citet{helly07}.
{\nbf Note that since the merger tree algorithm described above  
only makes use of equation~(\ref{eq:dnp_mod}) for
progenitor masses $M_1<M_2/2$ it is only the behaviour of
$G(\sigma_1/\sigma_2,\delta_2/\sigma_2)$ for $M_1<M_2/2$ 
($\sigma_1>\sigma_2$) that is constrained by comparison to
the Millennium Simulation merger trees. Consequently the
predictions of equation~(\ref{eq:dnp_mod}) for $M_1>M_2/2$ 
are of no relevance.
}

\label{sec:cmf}
\begin{figure*}
\includegraphics[width=16.5cm]{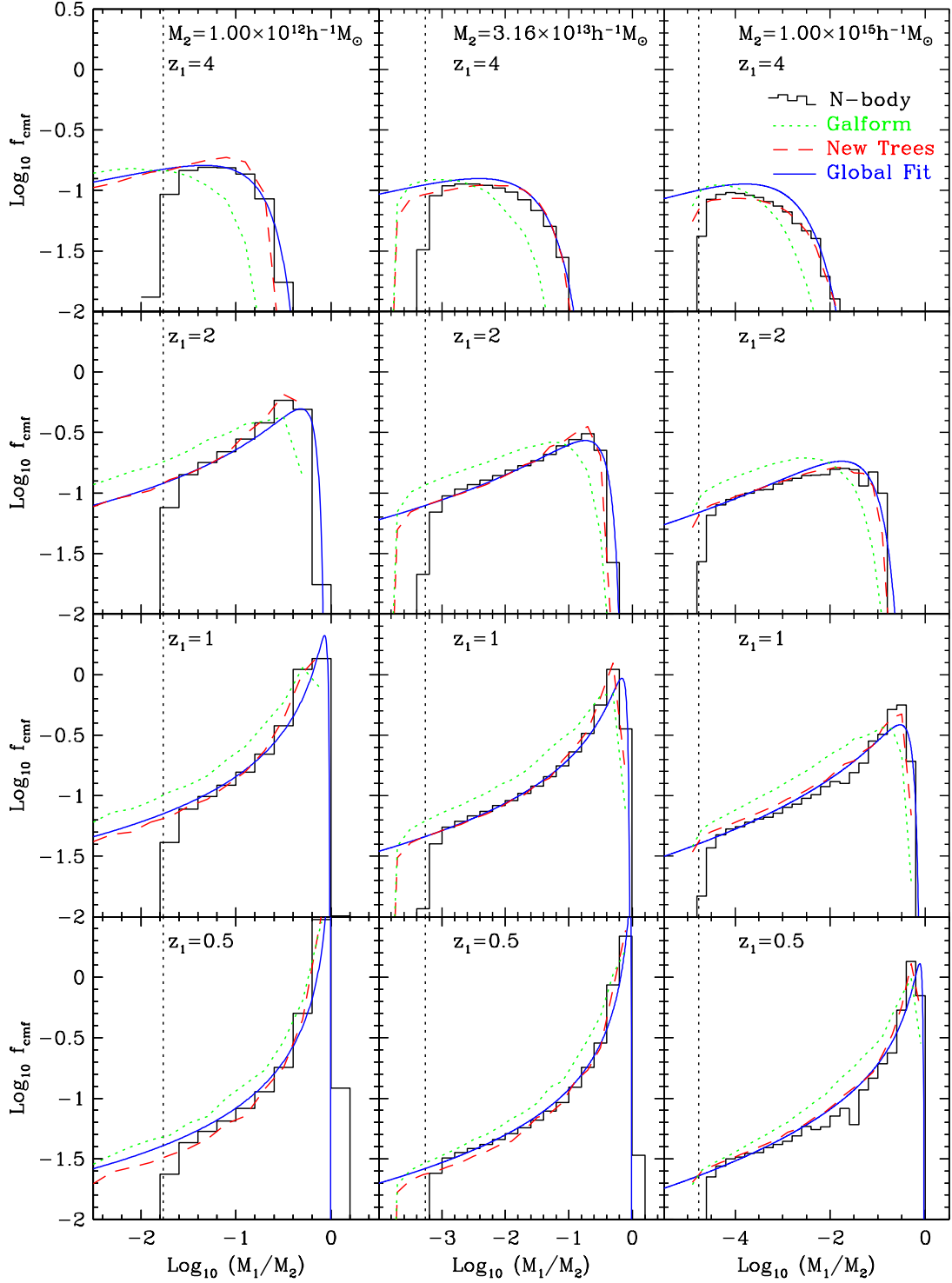}
\vskip -0.8cm
\caption{ The fraction of mass in progenitor halos of mass $M_1$ in
bins of $\log_{10} M_1/M_2$ at redshifts $z_1=0.5, 1, 2$ and~$4$ as
indicated, for three different masses $M_2$ (indicated at the top of
each column).  The histograms show the results from the Millennium
Simulation while the dotted and dashed curves are the corresponding
conditional mass functions given by the original GALFORM Monte-Carlo
algorithm and our new modified algorithm respectively. The solid curve
shows an analytic fit to the whole set of conditional mass functions
as described in \citet{helly07}.  The vertical dotted line indicates
the 20~particle mass resolution of the Millennium simulation.  }
\label{fig:cmf}
\end{figure*}

To simplify the problem still further we make the assumption that
\begin{equation}
G(\sigma_1/\sigma_2,\delta_2/\sigma_2) = G_0  \
\left( \frac{\sigma_1}{\sigma_2} \right)^{\gamma_1} \
\left( \frac{\delta_2}{\sigma_2} \right)^{\gamma_2} ,
\end{equation}
which can be considered as a first order Taylor series approximation for
$\ln G$ in terms of $\ln(\sigma_1/\sigma_2)$ and $\ln
(\delta_2/\sigma_2)$. 
This functional form is particularly convenient. The two terms
$G_0$ and $\left( {\delta_2}/{\sigma_2} \right)^{\gamma_2}$ have no
dependence on $M_1$ and so just enter the integrals in equations
(\ref{eq:P}) and~(\ref{eq:F}) as multiplicative constants. 
The term $\left( {\sigma_1}/{\sigma_2} \right)^{\gamma_1}$ alters
the distribution $dN/dM_1$ and the
integrands in both (\ref{eq:P}) and~(\ref{eq:F}), but has simple
analytic properties that allow a very fast implementation of the
splitting algorithm (see Appendix~\ref{appendix}).

\section{Comparison with the Millennium Simulation}
\label{sec:compare}

The Millennium Simulation \citep[MS,][]{springel05} is, to date, the
largest N-body simulation of a cosmologically representative
volume. It uses $N=2160^3$ particles in a comoving cube of side $L=500
h^{-1}$~Mpc to follow the non-linear gravitational evolution of a
Gaussian random density field drawn from a power spectrum consistent
with cosmological constraints from 2dFGRS \citep{percival01} and the
first year WMAP data \citep{spergel03}. The cosmological density
parameters are $\Omega_{\rm m}=0.25$, $\Omega_{\rm b}=0.045$ and
$\Omega_{\Lambda}=0.75$, the Hubble parameter
$h=H_0/100$~km~s$^{-1}$~Mpc$^{-1}=0.73$ and the linear amplitude of the
density fluctuations in spheres of radius $8h^{-1}$~Mpc is
$\sigma_8=0.9$. At each of over $60$ output times a catalogue
of friends-of-friends \citep{davis85} groups was constructed and
the descendant of each group found at the subsequent timestep.
Details of the construction of these merger trees and 
and their statistical properties can be found in \citet{helly07}.
Below we compare a variety of the statistics they estimated with
the results of the original \citet{cole00} GALFORM MC algorithm
and our new modified algorithm.

\subsection{Conditional Mass Functions}
\label{sec:cmf}

In Fig.~\ref{fig:cmf} we compare the conditional mass functions of the
MS merger trees with those of the MC algorithms. Here for halos of
various masses $M_2$ at redshift $z_2=0$ we find what fraction of
their mass is in progenitor halos of mass $M_1$ at various earlier
redshifts $z_1$. The histograms show the results of the MS while the
solid curves show an analytic fit described in \citet{helly07}. The
results of the original GALFORM algorithm are shown by the dotted
curves.  
As has been noted by \citet{helly07} these conditional mass
functions evolve more rapidly than those of the simulation. Thus, the
``GALFORM 2000'' algorithm strongly underpredicts the number of high mass
progenitors at high redshift. 
{\nbf That the EPS theory gives
predictions that evolve more rapidly with redshift than is found in N-body 
simulations has been noted previously
\citep[e.g][]{vdbosch02,wechsler02,lin03}. 
\citet{giocoli07} have shown that average halo formation times agree
better with the elliptical collapse model of \citet{smt01} than with
the spherical collapse EPS formalism. Furthermore, \citet{giocoli07} find that
scaling the time variable in EPS theory by the factor
$\sqrt{q}=\sqrt{0.707}=0.84$ that comes from fitting the elliptical
collapse model to N-body data results in formation time predictions
that better match the N-body data for a wide range of final masses
and redshifts. By reference to equation~(\ref{eq:dnp}) it can be seen
that our factor $G(\sigma_1/\sigma_2,\delta_2/\sigma_2)$ of
equation~(\ref{eq:dnp_mod}) can be viewed as a modification to 
the timestep $dz_1$. Thus, the elliptical collapse modelling 
of \cite{giocoli07} suggests that we should find
$G(\sigma_1/\sigma_2,\delta_2/\sigma_2) \approx 0.84$.
In fact, we expect a somewhat lower value as the Monte Carlo trees
of the original GALFORM algorithm evolve even more rapdily than
the analytic predictions of the EPS formalism
\citep{helly07}.
}

To find the best fit parameters we have minimised the rms difference
\begin{equation}
\sigma_{\rm cmf} = \left\langle \left( \log_{10} f_{\rm cmf}^{\rm MS}
(M_1|M_2)
-\log_{10} f_{\rm cmf}^{\rm MC}(M_1|M_2)
\right)^2
\right\rangle^{1/2}
\end{equation}
between MS data and the results of the MC algorithm over all twelve
panels in Fig.~\ref{fig:cmf}. With the exception of the lowest two
mass bins plotted in each panel, which were discarded as they are
influenced by the mass resolution of the MS, equal weight was given to
each bin. Initially we kept $\gamma_1=\gamma_2=0$ fixed and allowed only $G_0$
to vary. For $G_0=1$, the original GALFORM algorithm, the rms
difference $\sigma_{\rm cmf}=0.27$. 
{\nbf The best fitting value of $G_0$ is $0.79$,
as anticipated, somewhat smaller than the factor $0.84$  from
\cite{giocoli07}, and this reduces the rms
difference significantly to $\sigma_{\rm cmf}=0.12$. }
However, this is a compromise and the data from the different panels of 
Fig.~\ref{fig:cmf} prefer different values of $G_0$.
This can be accommodated by allowing $\gamma_1$ and
$\gamma_2$ to vary. As $\delta_2/\sigma_2$ is an increasing function
of the final mass $M_2$, a positive $\gamma_2$ would give a relatively
higher merger rate for high mass, $M>M_*$, halos (where the
character mass, $M_*$, has the usual definition of $\sigma(M_*)=\delta$).
Choosing $\gamma_1>0$ skews the shape of the progenitor mass
functions by boosting the ratio of low mass to high mass
progenitors. Since $\sigma_1>\sigma_2$, setting $\gamma_1>0$ boosts
the overall merger rate and so needs to be compensated for by a the
lower value of $G_0$.  Allowing all three parameters to vary,
consistently good fits are found with $G_0=0.57$, $\gamma_1=0.38$ and
$\gamma_2=-0.01$ with a reduced rms deviation from the MS data
of $\sigma_{\rm cmf}=0.055$. Over most of the range over which it is
employed $G(\sigma_1/\sigma_2, \delta_2/\sigma_2)$ remains less than but
of order unity. The conditional mass functions produced by this new
set of trees are shown by the dashed lines in Fig.~\ref{fig:cmf}. We
see that this minor change to the merger tree algorithm has resulted
in merger trees that are in good agreement with the N-body simulation
results over a wide range of masses and redshifts. The only mass bins
where the MC and MS conditional mass functions are not in good agreement
are the bins with $M_1>M_2$. In a truly hierarchical model such as is
produced by our algorithm $M_1$ can never be greater than $M_2$. In the MS
data \citet{helly07} noted that $M_1>M_2$ happens occasionally for low mass
halos due to the temporary, premature linking of FOF groups.

\begin{figure*}
\includegraphics[width=16.5cm]{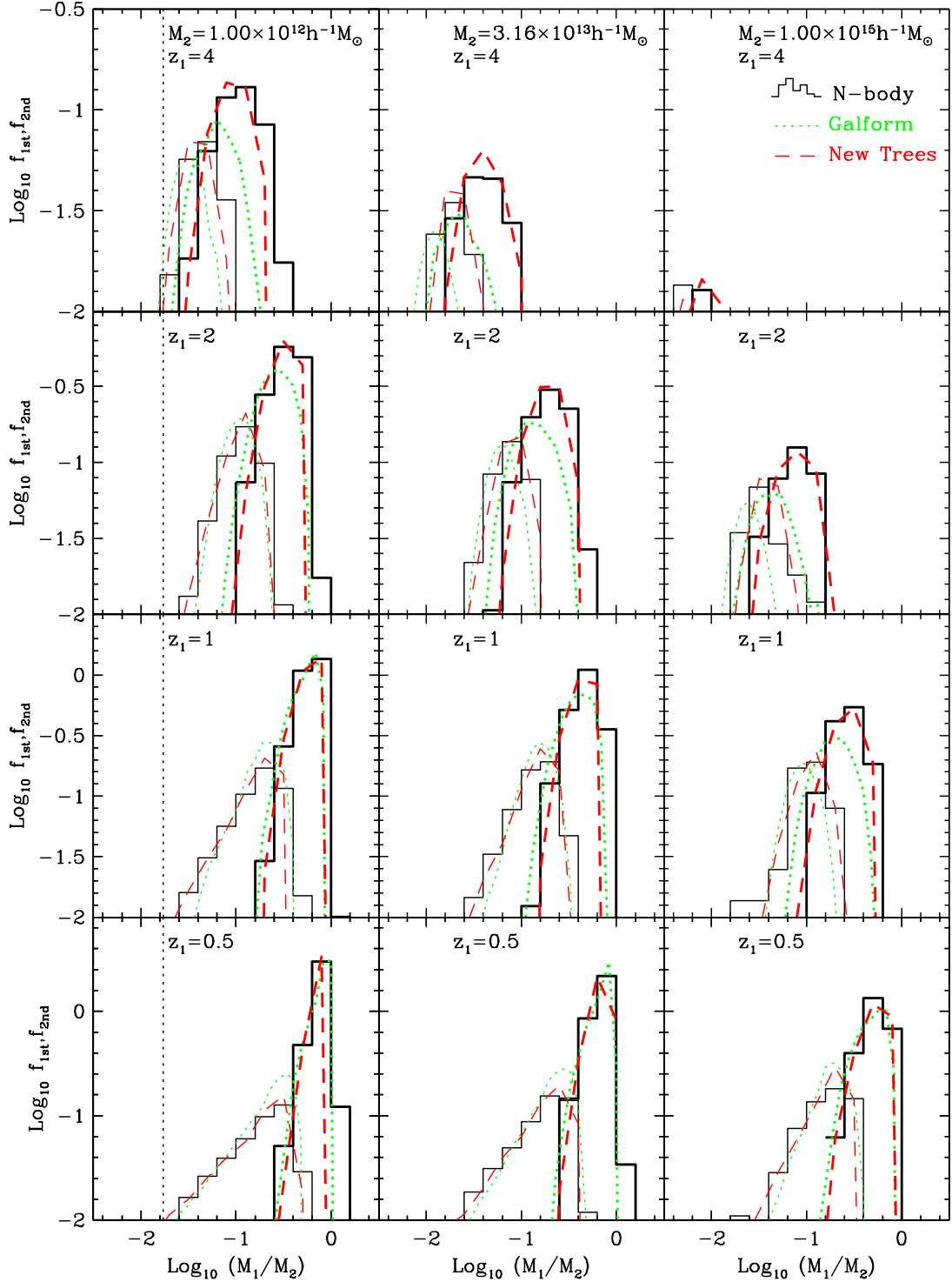}
\vskip -1cm
\caption{The mass distributions of the first and second most massive
progenitors. The plotted quantities $f_{\rm 1st}$ and $f_{\rm 2nd}$
are the contributions to the overall conditional mass functions
plotted in Fig.~\ref{fig:cmf} provided by the 1st and 2nd most massive
progenitors respectively. The panels correspond directly to those
of Fig.~\ref{fig:cmf} and are labelled by the final 
halo mass $M_2$ and redshift $z_1$ of the progenitors.
The histograms show the results from the Millennium Simulation
with distribution the $f_{\rm 1st}$ plotted with heavy lines and
$f_{\rm 2nd}$ with light lines. The corresponding predictions of
the GALFORM and new Monte-Carlo algorithms are shown by the heavy 
($f_{\rm 1st}$) and light ($f_{\rm 2nd}$) 
dotted and dashed curves respectively.
The $20$ particle mass resolution of the Millennium Simulation is
shown by the vertical dotted line, but only plays a role
for the $z=4$ progenitors of the lowest mass, $M_2=10^{12}$~\Msol, halos.
}
\label{fig:1st2nd}
\end{figure*}

\subsection{Main Progenitor Mass Functions}

The parameters of our new MC merger tree algorithm were tuned to
produce good agreement with the conditional mass functions plotted in
Fig.~\ref{fig:cmf} and so the level of agreement is perhaps not
surprising. However we can go further and test the success of the
algorithm by testing other statistical properties of the merger trees.
Fig~\ref{fig:1st2nd} plots the mass functions for the first and second
most massive progenitors for the same selection of redshifts and final
masses as in Fig~\ref{fig:cmf}. This is an interesting test of the 
merger trees as often in galaxy formation applications it is the most
massive progenitors and mergers between them that are most important
in determining the properties of the galaxies hosted by the halos. It
also tests an aspect of the merger trees that cannot be predicted 
by EPS theory alone as it involves how often one has a progenitor  of
a given mass and not just the mean number of such progenitors.

As noted in \citet{helly07} the mass functions for the first and
second most massive progenitors given by the original GALFORM algorithm
do a good job of matching the shape and relative positions of the two 
distributions, but systematically underestimate both masses with 
increasing redshift. This is completely remedied in the new algorithm
which matches the positions and shapes of the N-body distributions
extremely accurately.

\subsection{Major Mergers}

\begin{figure}
\includegraphics[width=8.5cm]{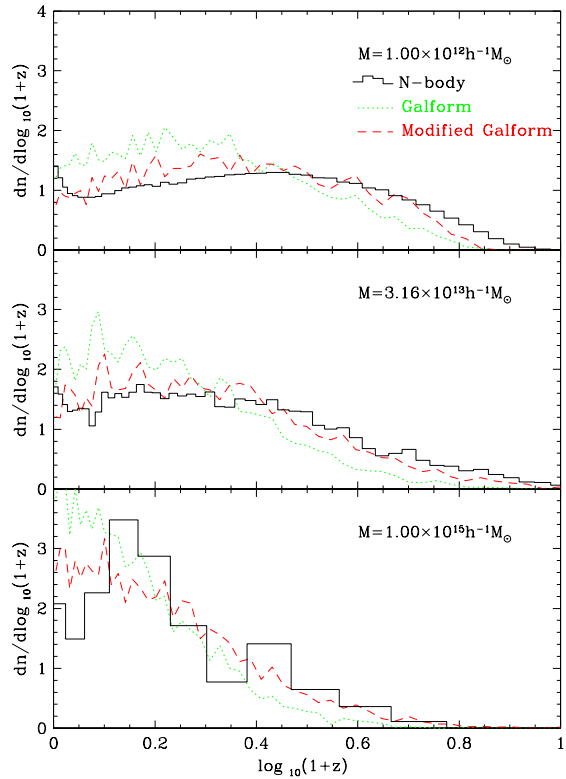}
\caption{The redshift distribution of the most recent major mergers
of halos with different final masses $M_2$. Here a major merger is
defined as a merger of the most massive progenitor of the final halo
with a second halo whose mass is at least $f_{\rm major}=0.3$ times
that of the main progenitor. The histogram shows the N-body results
and the dotted and dashed lines show the results from the original GALFORM
and new Monte-Carlo algorithm respectively.
}
\label{fig:majmerg}
\end{figure}

Another important aspect of the merger trees is the occurrence of major 
mergers. In galaxy formation models major mergers between galaxies,
which occur after halo mergers, are often deemed to be responsible for
initiating bursts of star formation and for converting disc galaxies
to spheroidal systems. Thus it is interesting to see what level of 
agreement our new algorithm has with estimates from the MS.
Fig.~\ref{fig:majmerg} compares the redshift distribution of the
most recent major merger for halos of various final masses. Here
a major merger has been defined as a merger between two halos where the
smaller is at least fraction $f_{\rm major}=0.3$ of the mass of the
larger. At each redshift we find the most massive progenitor of the
final halo and record the lowest redshift at which one of these
progenitors is undergoing a major merger.

It was found in \citet{helly07} that the original GALFORM algorithm,
shown by the dotted line in Fig.~\ref{fig:majmerg}, significantly
overestimated the number of recent major mergers.
Fig.~\ref{fig:majmerg} shows that this shortcoming is very largely
overcome by our new algorithm. The redshift distributions of the most
recent major mergers match accurately the overall shape of those from
the MS including their dependence on final halo mass.

\begin{figure}
\includegraphics[width=8.5cm]{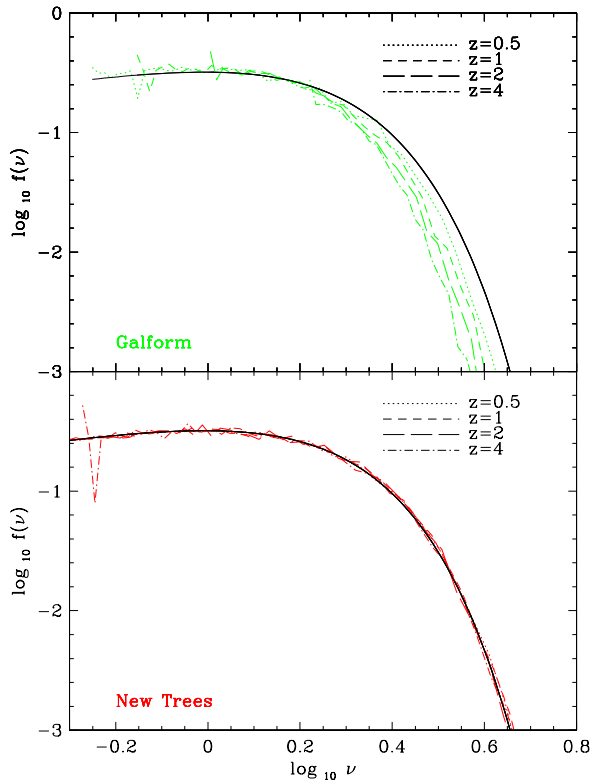}
\caption{The solid curve shows the Sheth-Tormen halo mass function
expressed in terms of the variable $\nu=\delta(z)/\sigma(M)$. This
is compared to mass functions at redshifts $z=0.5$,$1$,$2$ and $4$
determined by generating grids of merger trees starting at $z=0$
and counting their progenitors at these higher redshifts. The top
panel shows the results for the original GALFORM merger trees and the
bottom panel the results for our new merger trees.
}
\label{fig:tmf}
\end{figure}

\subsection{Overall Mass Functions}

The above comparisons to the results of the MS test the mass range
of the merger trees that is most important for galaxy formation 
applications, but it is also interesting to probe whether our new
algorithm remains plausible for much larger ranges in mass.
{\nbf \citet{st99} \citep[see also][]{jenkins01} }
have shown that for a wide range of initial
conditions and redshifts that the halo mass function has a universal
form. A good analytic match to this universal form for the fraction of
mass in  halos of mass $M$ is provided by the \citet{st02} mass function
\begin{equation}
f(M)\, d\ln M = f_{\rm ST}(\nu) 
\left\vert \frac{d\ln\nu}{d\ln M} \right\vert \,
d\ln M ,
\label{eq:mf}
\end{equation}
where
\begin{equation}
f_{\rm ST}(\nu) = A \sqrt{\frac{2 a}{\pi}} 
\left[ 1+ \left( \frac{1}{a\nu^2}\right)^p \right]
\nu \exp(-a \nu^2/2)
\label{eq:st}
\end{equation}
with $A=0.322$, $a=0.707$, $p=0.3$ and the mass dependent
variable $\nu=\delta/\sigma(M)$. Taking this as a good description
of the mass distribution of halos at redshift $z=0$ one can
generate a grid of merger trees rooted at $z=0$, weight them
by their redshift $z=0$ abundance and compute the overall
abundance of progenitor halos at any earlier redshift. 
If the merger the tree algorithm is in good agreement with N-body simulations
then these $z>0$ mass functions should be in good agreement with the
Sheth-Tormen mass function evaluated at that redshift.

Fig.~\ref{fig:tmf} compares the Sheth-Tormen mass function with those
determined with both the original GALFORM and new merger trees. In the
top panel one sees that the high mass exponential cut off to the
mass function systematically moves to lower $\nu$ at higher redshift.
In other words, as we saw with the conditional mass functions in
Fig.~\ref{fig:cmf}, the characteristic mass evolves too rapidly in these
trees. In contrast in the lower panel we see that with our new trees
this systematic error is greatly reduced and the abundance of high mass
halos matches the Sheth-Tormen prediction quite accurately over a wide
range of redshift.
Note that for each merger tree mass function the turnover at low
masses (low $\nu$) is
due to the specified mass resolution of the trees. At higher redshift
a fixed mass implies higher $\nu=\delta(z)/\sigma(M)$ and so the mass
resolution causes deviations at higher and higher $\nu$.

\section{Discussion}
\label{sec:disc}

{\nbf It is perhaps surprising that an algorithm motivated by EPS
theory, which is only a function of the smoothed linear theory
overdensity at a point, is able to accurately describe the complete
merger histories of dark matter haloes in a fully non-linear N-body
simulation.  The EPS theory, as derived by \cite{bcek}, makes the
following series of assumptions none of which can rigorously be true.
It assumes that virialized halos form when the linear theory
overdensity equals the threshold for collapse given by the pure
spherical collapse model; the linear overdensity at a given point in
space is assumed to vary with the smoothing scale as an uncorrelated
(Brownian) random walk (the sharp $k$-space filtering approximation);
when assigning mass points to halos of mass $M$ no condition is set to
require that these mass points should lie in spatially localised
regions capable of forming halos of that mass.  

One might have thought
that a more natural starting point for developing an accurate merger
tree algorithm would have been the ellipsoidal collapse model of
\cite{st99,smt01} as its mass function much more accurately matches that of
N-body simulations (although a free parameter, $q$\footnote{This
parameter is denoted $a$ in \cite{st99} and $q$ in \cite{giocoli07}}, is
adjusted to achieve this fit). However, it is not easy to work with
this model as there is no simple analytic expression for the
conditional mass function for small timesteps.  Furthermore, the
results of this more complicated model can often be approximated by
minor modifications of the formulae that are derived using the EPS
formalism. For example, \cite{giocoli07} have shown the inserting a
factor of $\sqrt{q}=0.84$ into the EPS formation time formula of
\cite{lc93} results in a reasonable match to the predictions of the
ellipsoidal collapse model.  Our algorithm was motivated by the EPS
formalism, but the modification we introduce in
equation~(\ref{eq:dnp_mod}) means that its predictions are no longer
those of the EPS formalism.  If instead one were trying to come up
with an algorithm based on the ellipsoidal collapse model, then the end
result might well be very similar. In fact, as noted in
Section~\ref{sec:cmf}, the $\sqrt{q}=0.84$ factor advocated by
\cite{giocoli07} is equivalent to our $G_0$ factor.  The other
assumptions of the EPS theory, listed above, and not addressed in the
ellipsoidal collapse model must also play a role in determining merger
histories. By adopting the modification defined in
equation~(\ref{eq:dnp_mod}) and fitting directly to N-body simulation
results, our model is fitting the net effect of all the additional
physics and not just that due to the shape of the density
perturbation.

 After we completed this project \cite{neistein07} presented an
alternative algorithm to generate dark matter halo merger trees based
on fitting log-normal distributions to progenitor mass functions
expressed in scaled mass and time variables.  Their algorithm, which
is of very comparable speed to ours, is also tuned to fit the conditional mass
functions of merger trees extracted from the Millennium
simulation. There will be some differences in the results of the two
algorithms because the Millennium Simulation merger trees used by
\cite{neistein07} are not the simple friends-of-friends merger trees
we constructed for this project, but instead the ``{\sc DHALO}'' merger
trees that were constructed by the Durham Group and used in the
semi-analytic galaxy formation model of \cite{bower06}.  Both sets of
trees are based on the same catalogues of friends-of-friends groups,
but the ``{\sc DHALO}'' algorithm uses additional information concerning
substructures identified using {\sc SUBFIND} \citep{springel01}.  (There is
some discussion of the additional criteria useful for galaxy formation
calculations in \cite{harker06}.)  We
opted not to use these trees since a criterion that delays the time at
which the merger is deemed to take place has the side effect of
causing some halos to loose mass prior to the merger.  This
artificially increases the occurence of progenitor halos that are more
massive than their descendents and so slightly distorts the 
conditional mass functions. }

\section{Conclusions}
\label{sec:conc}

We have presented a new Monte-Carlo algorithm to generate dark matter
halo merger trees. The algorithm is a modification of the Extended
Press-Schechter algorithm described in \citet{cole00}. The change we
have made to the algorithm {\nbf was} motivated empirically and tuned to match
the conditional  mass function of halos extracted from the Millennium
Simulation \citep[MS,][]{springel05}. We find that not only can we get a very
accurate match to these conditional mass functions over a wide range
of mass and redshift, but that the other statistical properties of the
new trees match very well those from the Millennium Simulation. The
improvement in accuracy over the algorithm previously used in the
GALFORM semi-analytic code \citet{cole00} is very significant and
should make the new algorithm a very useful tool.

While our algorithm has been tuned to match the results of MS,
which is a particular $\Lambda$CDM model, we would expect it to a 
significant improvement over EPS based algorithms for quite a wide 
range of CDM-like initial conditions. The overly rapid evolution
in the typical mass of progenitor halos
was a generic problem with the old algorithm and the reduced merger
rate of the new algorithm should be an improvement in all cases.
We have made a fortran90 implementation of algorithm available
at http://star-www.dur.ac.uk/$\tilde{\hphantom{n}}$cole/merger\_trees .

\section*{Acknowledgements}
{\nbf We thank the referee, Ravi Sheth, for comments
that improved the paper.} We thank Yu Lu for finding an error in our code
in time for us to remedy the code, correct the published version 
of the paper and make improvements to the appendix.
The {\nbf Millennium S}imulation used in this paper was
carried out as part of the programme of the Virgo Consortium on the
Regatta supercomputer of the Computing Centre of the
Max-Planck-Society in Garching. {\nbf 
Data for the halo population in this simulation, as well
as for the galaxies produced by several different galaxy formation
models, are publically available at http://www.mpa-garching.mpg.de/millennium
and under the ``downloads'' button at http://www.virgo.dur.ac.uk/new .}
This work was supported in part by the 
PPARC rolling grant for Extragalactic and cosmology research at Durham.

\setlength{\bibhang}{2.0em}

\appendix
\section{The Split Algorithm}
\label{appendix}

Given a halo of mass $M_2$ at redshift $z$, the task of the split
algorithm is to take a small step, $\Delta z$, to higher redshift
determine the mass accreted in this interval in unresolved halos with
masses less than $M_{\rm res}$ and determine whether or not the halo
undergoes a binary split.  If the halo does split then it must
determine the masses of the two fragments.
In the description below we make use of the following notation.
We denote minus logarithmic slope of the $\sigma(M)$ relation by
$\alpha(M)=-d\ln\sigma/d\ln M$ and its values at masses $M_2$,
$M_2/2$ and $M_1=qM_2$ by $\alpha_2$, $\alphah$ and $\alpha_1(q)$
respectively. Similarly we denote the values of $\sigma(M)$ at
$M_2$, $M_2/2$, $M_{\rm res}=q_{\rm res}M_2$ and $M_1$ by 
$\sigma_2$, $\sigmah$, $\sigma_{\rm res}$ and $\sigma_1(q)$ respectively.
With this notation the expression in equation~(\ref{eq:dnp}) for
number of fragments produced per unit interval of $q$ produced
in a redshift step $\Delta z$ can be written as
\begin{equation}
\frac{dN}{dq} = S(q) \ R(q) \ \Delta z ,
\label{eq:dnq}
\end{equation}
where
\begin{equation}
S(q) = \sqrt{\frac{2}{\pi}} \ B\, \alphah\, q^{\eta-1}   \
\frac{G_0}{2^{\mu\gamma_1}} \left(\frac{\delta}{\sigma_2}\right)^{\gamma_2} \,
 \left(\frac{\sigmah}{\sigma_2} \right)^{\gamma_1}\,
\frac{d\delta}{dz},
\end{equation}
\begin{equation}
R(q) = 
\frac{\alpha_1(q)}{\alphah} \
\frac{V(q)}{B q^\beta} \
\left( \frac{(2q)^\mu \sigma_1(q)}{\sigmah}\right)^{\gamma_1}  
\label{eq:R}
\end{equation}
\begin{equation}
V(q) = \frac{\sigma_1(q)^2}{[\sigma_1(q)^2-\sigma_2^2]^{3/2}} ,
\end{equation}
and $\eta=\beta-1-\gamma_1\mu$.
We have written the expression for $dN/dq$ in this form so that,
as detailed below, we can choose the parameters $B$, 
$\beta$ and $\mu$ such that $R(q) < 1$ for $q_{\rm res}<q<1/2$ and 
$S(q)\propto q^{\eta-1}$ is a simple power law.
This results in several very useful properties. First, 
\begin{equation}
  N_{\rm upper}= \int_{q_{\rm res}}^{1/2} S(q) \, dq \ \Delta z 
  \label{eq:dz}
\end{equation}
provides an upper limit on the expected number of resolved fragments
split off the main halo in step $\Delta z$. We use this to choose the
step size by taking $\Delta z$ to be the minimum 
of $\epsilon_1 \sqrt{2} (\sigmah^2-\sigma_2^2)^{1/2}/ d\delta/dz$
and the value given by equation~(\ref{eq:dz}) when
$N_{\rm upper}= \epsilon_2$ (by default, we take $\epsilon_1=\epsilon_2=0.1$).
The first constraint ensures that the exponent in
equation~(\ref{eq:cmff}) is small so that the equation~(\ref{eq:lin}) 
is correct to first order and the second constraint ensures that
multiple splittings in one timestep are negligible.

Having determined $\Delta z$, the next step is to evaluate $F$ from
equation~(\ref{eq:F}) to determine fraction of mass that is accreted in
unresolved halos in this timestep. The expression defining $F$
can be simplified to the form
\begin{equation}
F=  \sqrt{\frac{2}{\pi}} J(u_{\rm res})\ \frac{G_0}{\sigma_2} \left(\frac{\delta_2}{\sigma_2}\right)^{\gamma_2} \
\ \frac{d\delta}{dz} \, \Delta z,
\end{equation}
where we have made the substitution
$u=\sigma_2/(\sigma_1^2-\sigma_2^2)^{1/2}$ and the integral
\begin{equation}
J(u_{\rm res}) = \int_0^{u_{\rm res}} (1+1/u^2)^{\gamma_1/2} du,
\end{equation}
with $u_{\rm res}=\sigma_2/(\sigma_{\rm res}^2-\sigma_2^2)^{1/2}$.
Since this integral has no dependence on $M_2$, $z$ or $\sigma(M)$
it can be tabulated as a simple look-up table for any chosen
value of the parameter $\gamma_1$. In the original GALFORM
algorithm it reduces to $J(u_{\rm res}) = u_{\rm res}$.

The next step is to generate the first of three uniform
random variables in the range $0$ to $1$. If this first
variable, $r_1$, is greater than $N_{\rm upper}$ evaluated
with the selected $\Delta z$ then no split
occurs at this timestep and $M_2$ is just reduced to 
$M_2(1-F)$ to account for the accreted mass.  
If $r_1<N_{\rm upper}$ we generate a second random variable $r_2$
and transform it using $q= (q_{\rm res}^\eta +(2^{-\eta}-q_{\rm res}^\eta) r_2)^{1/\eta}$
so that it is drawn  from the power-law distribution $q^{\eta-1}$ in
the range $q_{\rm res}<q<1/2$.  Finally we generate a third
random variate $r_3$ and only accept $q$ if $r_3<R(q)$. In the
case that $q$ is rejected we again simply reduce $M_2$ to
$M_2(1-F)$, but if $q$ is accepted we generate two fragments
with masses $qM_2$ and $M_2(1-F-q)$. This rejection step ensures
that $q$ is being drawn with the correct normalization
from the probability distribution defined by equation~(\ref{eq:dnq}).

For this algorithm to work we require $R(q)<1$ for 
$q_{\rm res}<q<1/2$. Referring to equation~(\ref{eq:R}), in all CDM models,
$\alpha(M)>0$ and $d(\alpha)/dM>0$ and so the first term
${\alpha_1(q)}/{\alphah}$ is necessarily less than one.
Also these conditions imply that the function $V(q)$
is monotonically increasing and $\ln(V)$ versus $\ln(q)$ is concave upwards 
for $0<q<1/2$. This means that
$V(q)$ is bounded from above by the power law $Bq^\beta$
chosen to satisfy $B q^\beta = V(q)$ for $q=q_{\rm res}$ and $q=1/2$.
In other words with this choice of $B$ and $\beta$ the second term
in equation~(\ref{eq:R}),
$V(q)/Bq^\beta$, is less than or equal to one. 
Finally if we choose
\begin{equation}
\mu= \cases{ \alphah  & if \quad $\gamma_1>0$  \cr
 \displaystyle{-\frac{\ln(\sigma_{\rm res}/\sigmah)}{\ln{2q_{\rm
 res}}}}& if \quad $\gamma_1<0$  }
\end{equation}
then regardless of the sign of $\gamma_1$ the last factor
$\left( \frac{(2q)^\mu \sigma_1(q)}{\sigmah}\right)^{\gamma_1}$,
is also less than or equal to one and so $R(q)$ is always less than
one as required.

The merger tree produced by this algorithm has no directly imposed 
time/redshift  resolution and comprises of only binary mergers.
However, we typically rebin each merger tree onto a discrete
grid of predefined redshift snapshots. With this coarser time
resolution the mergers occurring between snapshots can involve three
or even more halos.


\begin{thebibliography}{}

\bibitem[Benson et al.(2005)]{benson05} 
Benson, A.~J., Kamionkowski, M.,  Hassani, S.~H.\ 2005, \mnras, 357, 847 

\bibitem[Bond et al.(1991)]{bcek} 
Bond, J.~R., Cole, S., Efstathiou, G.,  Kaiser, N.\ 1991, \apj, 379, 440 

\bibitem[Bower(1991)]{bower91} 
Bower, R.~G.\ 1991, \mnras, 248, 332

\bibitem[Bower et al.(2006)]{bower06} 
Bower, R.~G., Benson, A.~J., Malbon, R., Helly, J.~C., Frenk, C.~S., 
Baugh, C.~M., Cole, S., \& Lacey, C.~G.\ 2006, \mnras, 370, 645 
 
\bibitem[Cole(1991)]{cole91} 
Cole, S.\ 1991, \apj, 367, 45 

\bibitem[Cole et al.(2000)]{cole00}
Cole, S., Lacey, C.~G., Baugh, C.~M., \& Frenk, C.~S.\ 2000, \mnras, 319, 168 

\bibitem[Cole et al.(2007)]{helly07} 
Cole, S., Helly, J.C.,  Frenk, C.S., Parkinson, H.,
\ 2007, \mnras, accepted.

\bibitem[Davis et al.(1985)]{davis85} 
Davis, M., Efstathiou, G., Frenk, C.~S.,  White, S.~D.~M.\ 1985,
\apj, 292, 371 

\bibitem[Efstathiou et al.(1988)]{efstathiou88} 
Efstathiou, G., Frenk, C.~S., White, S.~D.~M.,  Davis, M.\ 1988,
\mnras, 235, 715 

\bibitem[Eke et al.(1996)]{ecf96} 
Eke, V.~R., Cole, S.,  Frenk, C.~S.\ 1996, \mnras, 282, 263 

\bibitem[Jenkins et al.(2001)]{jenkins01} 
Jenkins, A., Frenk, C.~S., White, S.~D.~M., Colberg, J.~M., Cole, S.,
Evrard, A.~E., Couchman, H.~M.~P.,  Yoshida, N.\ 2001, \mnras, 321,
372 

\bibitem[Giocoli et al.(2007)]{giocoli07} 
{\nbf Giocoli, C., Moreno, J., Sheth, R.~K., \& Tormen, G.\ 2007,
  \mnras, 376, 977 }


\bibitem[Harker et al.(2006)]{harker06} 
{\nbf Harker, G., Cole, S., Helly, J., Frenk, C., \& Jenkins, A.\ 2006, \mnras, 367, 1039 
}

\bibitem[Kauffmann \& White(1993)]{kauffmann93} 
Kauffmann, G.,  White, S.~D.~M.\ 1993, \mnras, 261, 921 

\bibitem[Lacey  \& Cole(1993)]{lc93} 
Lacey, C.,  Cole, S.\ 1993, \mnras, 262, 627 

\bibitem[Lacey  \& Cole(1994)]{lc94} 
Lacey, C.,  Cole, S.\ 1994, \mnras, 271, 676 

\bibitem[Lin et al.(2003)]{lin03} 
{\nbf Lin, W.~P., Jing, Y.~P., \& Lin, L.\ 2003, \mnras, 344, 1327 }


\bibitem[Neistein \& Dekel(2007)]{neistein07} 
{\nbf Neistein, E., \& Dekel, A.\ 2007, MNRAS accepted, astro-ph/07081599
}
\bibitem[Percival et al.(2001)]{percival01}
Percival, W.~J., et al.\ 2001, \mnras, 327, 1297 

\bibitem[Press \& Schechter(1974)]{ps74} 
Press, W.~H., \& Schechter, P.\ 1974, \apj, 187, 425 

\bibitem[Sheth \& Lemson(1999)]{sheth99} 
Sheth, R.~K., \& Lemson, G.\ 1999, \mnras, 305, 946 

\bibitem[Sheth \& Tormen(1999)]{st99}
{\nbf Sheth, R.~K., \& Tormen, G.\ 1999, \mnras, 308, 119 }

\bibitem[Sheth et al.(2001)]{smt01} 
Sheth, R.~K., Mo, H.~J., \& Tormen, G.\ 2001, \mnras, 323, 1 

\bibitem[Sheth \& Tormen(2002)]{st02} 
Sheth, R.~K.,  Tormen, G.\ 2002, \mnras, 329, 61

\bibitem[Somerville \& Kolatt(1999)]{somerville99} 
Somerville, R.~S., \& Kolatt, T.~S.\ 1999, \mnras, 305, 1 

\bibitem[Spergel et al.(2003)]{spergel03} 
Spergel, D.~N., et al.\ 2003, \apjs, 148, 175 


\bibitem[Springel et al.(2001)]{springel01} 
{\nbf Springel, V., White, S.~D.~M., Tormen, G., \& Kauffmann, G.\ 2001, \mnras, 328, 726 
}

\bibitem[Springel et al.(2005)]{springel05} 
Springel, V., et al.\ 2005, \nat, 435, 629 


\bibitem[van den Bosch(2002)]{vdbosch02} 
{\nbf van den Bosch, F.~C.\ 2002, \mnras, 331, 98 }

\bibitem[Wechsler et al.(2002)]{wechsler02} 
{\nbf Wechsler, R.~H., Bullock, J.~S., Primack, J.~R., Kravtsov, A.~V., \& Dekel, A.\ 2002, \apj, 568, 52 }



\end{thebibliography}
\end{document}